\documentclass[12pt]{aa}

\usepackage{graphicx}
\usepackage[T1]{fontenc}
\usepackage{amsmath}
\usepackage[utf8]{inputenc}
\usepackage{txfonts}
\usepackage{hyperref}

\hypersetup{colorlinks=false}

\begin{document}

   \title{Effects of Alfvénicity on the compressible energy cascade in magnetohydrodynamic solar wind turbulence: Evidence from Parker Solar Probe, Solar Orbiter, and WIND observations}

   \author{Nahuel Andrés\inst{1,2} \and C.~A.~Gonzalez\inst{3} \and Norberto Romanelli\inst{4,5}}

   \institute{Universidad de Buenos Aires, Facultad de Ciencias Exactas y Naturales, Departamento de Física, Ciudad Universitaria, 1428 Buenos Aires, Argentina \email{nandres@df.uba.ar}
   \and CONICET -- Universidad de Buenos Aires, Instituto de Física Interdisciplinaria y Aplicada (INFINA), Ciudad Universitaria, 1428 Buenos Aires, Argentina
   \and Department of Physics, The University of Texas at Austin, Austin, TX, USA
   \and Department of Astronomy, University of Maryland, College Park, MD, USA
   \and Planetary Magnetospheres Laboratory, NASA Goddard Space Flight Center, Greenbelt, MD, USA}

   \date{Received \today; accepted \today}

\abstract
{The compressible magnetohydrodynamic (MHD) energy cascade in the solar wind is shaped by Alfvénicity, plasma fluctuations, and proton temperature, yet how they regulate the relative contributions of compressible and incompressible energy transfer across heliocentric distances remains poorly understood.}
{We investigate how Alfv\'enicity, plasma fluctuation amplitudes, and proton temperature regulate the compressible and incompressible MHD energy cascade rates in the solar wind, and how the two components of the compressible cascade, the Yaglom-like term and the purely compressible term, respond to changes in plasma conditions.}
{We apply compressible and incompressible exact relations to a large statistical sample of 2-hour intervals from WIND (1~au), Parker Solar Probe (PSP, $<0.25$~au), and Solar Orbiter (SolO, 0.3--1~au), classifying intervals as Alfv\'enic ($|\sigma_c|>0.75$) or non-Alfv\'enic ($|\sigma_c|<0.25$) based on the normalized cross helicity, and further separating by solar wind speed.}
{Velocity fluctuation amplitude is found to be the strongest predictor of the compressible cascade rate $\langle|\varepsilon_c|\rangle$ among the variables examined, followed by density and magnetic field fluctuations. $\langle|\varepsilon_c|\rangle$ and $\langle|\varepsilon_i|\rangle$ are tightly correlated across all regimes, with their ratio organized by Alfv\'enicity and wind speed: non-Alfv\'enic and Alfv\'enic fast winds cluster along $\langle|\varepsilon_c|\rangle = \langle|\varepsilon_i|\rangle$, while Alfv\'enic slow wind shows a systematic excess consistent with internal energy contributions. Across all regimes, the intervals with the largest $\langle|\varepsilon_c|\rangle$ are systematically the hotter ones, with increases of nearly 1.5 and two decades in Alfv\'enic slow and non-Alfv\'enic wind, respectively, a trend primarily carried by the purely compressible cascade, while the dominant Yaglom-like term appears to control the overall cascade magnitude.}
{Our results indicate that Alfv\'enicity and wind speed govern the efficiency of turbulent energy transfer across heliocentric distances. The tight coupling between compressible and incompressible cascade rates points toward the need for compressible, multi-fluid descriptions to fully capture the turbulent dissipation budget in the solar wind.}


\titlerunning{Alfvénicity and compressible MHD cascade in the solar wind}
\authorrunning{Andrés et al.}

\maketitle
\nolinenumbers

\section{Introduction}\label{sec:intro}

Magnetohydrodynamic (MHD) turbulence plays a central role in the dynamics of astrophysical and space plasmas, governing the transport of energy across a wide range of spatial and temporal scales and ultimately controlling plasma heating and dissipation \citep[e.g.,][]{M2011,V2019,Sa2020,Sc2022}. In the solar wind, MHD turbulence persists over a broad range of heliocentric distances and is characterized by fluctuations in velocity, magnetic field, and density spanning several orders of magnitude in scale \citep{BC2013,Chen2016}. These fluctuations interact nonlinearly, generating a turbulent cascade that transfers energy from large, energy-containing scales toward small kinetic scales, where dissipation and particle heating take place. A key quantity for characterizing this process is the energy cascade rate $\varepsilon$, which quantifies the flux of energy across scales. In incompressible MHD turbulence, the cascade rate can be directly estimated from in situ spacecraft measurements using exact relations derived from the MHD equations \citep{P1998b,P1998a}. These exact relations link third-order structure functions of plasma-field increments to the mean energy transfer rate, providing a powerful, model-independent diagnostic of turbulence \citep[e.g.,][]{SV2007,M2008,C2009,St2009,H2017a,H2017b,A2020,A2022,R2022,MSV2023,R2024a}.

However, the solar wind is not strictly incompressible. Density fluctuations, magnetic field magnitude variations, and compressive velocity components are ubiquitous, particularly in slow solar wind and near large-scale structures \citep[e.g.,][]{B1982,M1989,H2005,Chen2016}. This has motivated the development of exact relations for compressible turbulence models, which generalize the incompressible framework by explicitly incorporating compressible effects \citep{Ga2009,B2013,G2014,A2017b}. Recent formulations have demonstrated that the compressible cascade rate can differ substantially from its incompressible counterpart, both in amplitude and in its dependence on plasma conditions \citep{A2021,S2021,B2023,A2025}. In particular, density fluctuations have been shown to amplify the compressible cascade rate with the enhancement scaling with the turbulent Mach number and exhibiting a clear dependence on the thermal state of the plasma \citep{C2009,H2017a,A2021,A2025}. More specifically, \citet{B2023} showed a clear increase in the absolute value of both the compressible and incompressible cascade rates as heliocentric distance decreases, finding that isothermal and polytropic cascade rates exceed their incompressible counterpart as compressibility increases in the plasma. Determining when compressibility enhances, suppresses, or slightly modifies the cascade is therefore essential for a complete description of solar wind turbulence.

A key ingredient of solar wind turbulence is its degree of Alfvénicity, commonly quantified by the correlation between velocity and magnetic-field fluctuations or by the normalized cross helicity,
\begin{equation}\label{sigmac}
\sigma_c = \frac{\langle {\bf u}\cdot{\bf u}_{\rm A}\rangle}{\langle |{\bf u}|^2\rangle + \langle |{\bf u}_{\rm A}|^2\rangle},
\end{equation}
where ${\bf u}$ and ${\bf u}_{\rm A}\equiv{\bf b}/\sqrt{4\pi\rho}$ denote the bulk plasma and Alfv\'en velocity fluctuation vectors, respectively, and $\rho$ is the plasma mass density. The angular bracket refers to a time average over a specific window length. Alfvénic solar wind streams are characterized by outwardly propagating fluctuations that exhibit a strong alignment between velocity and magnetic field perturbations, resembling Alfvén waves \citep{B1971}. In contrast, non-Alfvénic intervals contain a more balanced proportion of counter-propagating modes and display weaker correlations between velocity and magnetic fluctuations, resulting in a richer mixture of plasma modes and enhanced intermittency \citep{B1971,BC2013,V2019,Si2022}. This distinction is closely related to the classical separation between fast and slow solar wind. Fast wind streams, originating from coronal holes, are typically highly Alfvénic, with large-amplitude, nearly incompressible fluctuations and relatively weak density variations. The slow Alfvénic solar wind, in contrast, is generally more compressible, and more intermittent, reflecting its more complex coronal origins and stronger interactions with large-scale structures \citep[e.g.,][]{DA2015,DA2022}. This picture is further expanded by the existence of a distinct population of Alfvénic slow wind that shares the high cross-helicity and low-compressibility of fast streams but at lower bulk speeds \citep{DA2021}. In particular, the slow Alfvénic wind exhibit kinetic and chemical composition properties that are distinct from fast solar wind \citep{d2019slow,alterman2025cross,huang2025temperature}, implying different source regions and acceleration process that breaks down the conventional fast-slow wind paradigm. Recent in situ observations from Parker Solar Probe (PSP) and Solar Orbiter (SolO) further show that Alfvénic turbulence dominates the inner heliosphere, even at relatively low wind speeds, whereas non-Alfvénic intervals become increasingly prevalent with distance from the Sun and within interaction regions \citep{Shi2021,DA2025}, with turbulence properties varying substantially among streams with similar speeds due to distinct coronal origins and large-scale structures \citep{Shi2021}. These observational results indicate that the impact of compressibility on MHD turbulence cannot be separated from the Alfvénic state of the plasma.

The dependence of the turbulent energy transfer rate on the degree of Alfvénicity has been investigated observationally for more than a decade. Using WIND observations at 1~au, \citet{smith2009} and \citet{St2010} showed that the incompressible cascade rate is strongly modulated by the normalized cross helicity, and that for $|\sigma_c| \gtrsim 0.75$ the transfer rate associated with the dominant outward-propagating component can become negative, indicating a back-transfer of energy toward large scales \citep[see also,][]{P2011}. In the high-latitude solar wind sampled by Ulysses, \citet{Ma2012} found that intervals with high cross helicity are associated with a reduced energy transfer rate, and that slow streams exhibit larger cascade rates than the neighboring fast streams, an enhancement attributed to the decorrelation between velocity and magnetic field fluctuations \citep[see also,][]{Ma2011}. More recently, \citet{R2024a} computed the signed incompressible cascade rate in the pristine solar wind upstream of Mars using MAVEN observations, finding negative median total cascade rates across the entire cross-helicity range, together with a power-law dependence of their magnitude on the total energy per unit mass of the fluctuations. While the median total cascade rate was found not to vary significantly with $\sigma_c$, the inward and outward pseudo-energy cascade rates do, with the most Alfvénic states associated with the most intense (negative) pseudo-energy rates. These authors attributed the prevalence of negative median rates at Mars to the comparatively low total fluctuation energy found there, following the result of \citet{St2010} that the range of $\sigma_c$ values associated with negative cascade rates expands as the total energy of the fluctuations decreases. A comprehensive account of these and related results is given in the recent review by \citet{MSV2023}. All these studies, however, rely on the incompressible exact relation and on signed transfer rates.

On theoretical grounds, density fluctuations enter the compressible exact laws through both the generalized Yaglom flux and through internal-energy terms, suggesting a direct amplification of $\varepsilon_c$ relative to $\varepsilon_i$ when compressibility is significant \citep{A2017b,A2021,S2021,A2025,J2025}. \citet{H2017a} showed that compressible cascade rates are enhanced in the slow solar wind, exhibiting a power-law scaling with the turbulent Mach number and a lower level of spatial anisotropy than in the fast wind. Yet, whether this amplification depends systematically on the relative contributions of density, velocity, and magnetic fluctuations, and how this hierarchy is modulated by the Alfvénic state of the turbulence, has not been addressed across different heliocentric distances. Moreover, the proton temperature has been found to correlate with the incompressible cascade rate at 1~au \citep{V1995,C2009,A2022}, but its relationship to $\varepsilon_c$ across different Alfvénicity and wind speed regimes remains poorly constrained. Here, we investigate the interplay between $\varepsilon_i$ and $\varepsilon_c$ in Alfvénic and non-Alfvénic solar wind turbulence using PSP and SolO in the inner heliosphere, complemented by WIND at 1~au, separating intervals by Alfvénicity and solar wind speed across heliocentric distances from $\sim$0.05~au to 1~au.

The paper is organized as follows. In Section \ref{sec:obs} we describe the spacecraft data and the event selection procedure. In Section \ref{sec:theory} we summarize the compressible and incompressible MHD exact relations used to estimate the energy cascade rates, and define the two compressible components. In Section \ref{sec:results} we present the statistical results: we first characterize the fluctuation amplitudes in Alfvénic and non-Alfvénic intervals, then examine the dependence of $\langle|\varepsilon_c|\rangle$ on plasma fluctuations and $\beta$, compare the compressible and incompressible cascade rates across wind speed and temperature regimes, and finally decompose the compressible flux into its two contributions. Section~\ref{sec:disc} discusses the physical interpretation of these results and places them in the context of existing theoretical and observational work.

\section{Observations and data selection}\label{sec:obs}

We investigated the energy cascade rate using in-situ observations from the WIND spacecraft at 1~au (2002--2020), employing magnetic field data from the Magnetic Field Investigation \citep[MFI;][]{lepping1995} and proton moments from the Three-Dimensional Plasma and Energetic Particle Investigation \citep[3DP;][]{lin1995}, with proton density from the Solar Wind Experiment \citep[SWE;][]{ogilvie1995swe} used as a quality reference. We incorporated PSP measurements \citep{Fox2016} from 2019 to 2024 (Encounters 3--21), restricted to heliocentric distances below 0.25~au where ion moments from the Solar Probe ANalyzer for Ions \citep[SPAN-I;][]{livi2022} are of highest quality; magnetic field data came from FIELDS \citep{bale2016} and electron densities from quasi-thermal noise \citep[QTN;][]{moncuquet2020} measurements. Solar Orbiter \citep[SolO;][]{Muller2020} data (2022--2024, 0.3--1~au) provided magnetic field from MAG \citep{horbury2020solar} and proton moments from the Solar Wind Analyzer--Proton and Alpha Sensor \citep[SWA-PAS;][]{owen2020solar}.

In this study, we estimate solar wind turbulence properties by analyzing 2-hour intervals of magnetic field and plasma moment data sampled at 5~s, based on the events identified by \citet{Go2026}. In particular, solar wind streams were categorized by their Alfvénicity, quantified by the normalized cross helicity (Eq.\,\ref{sigmac}). Alfvénic intervals were defined as those with $|\sigma_c| > 0.75$, while non-Alfvénic intervals satisfied $|\sigma_c| < 0.25$. The dataset used in this work contains 3806 Alfvénic and 633 non-Alfvénic intervals for WIND, 153 and 26 for SolO, and 591 and 16 for PSP, respectively. Additionally, we evaluated the sensitivity of our results to the choice of density measurements derived from PSP by comparing the main results obtained from SPAN-I ion densities with those derived from QTN, finding agreement within statistical uncertainties. Similarly, for WIND, we confirmed that the main statistical results are robust when using either 3DP or SWE density measurements. Finally, to ensure reliable estimates of the energy cascade rate at MHD scales, we retain only intervals in which the mean compressible cascade rate $\langle|\varepsilon_c|\rangle$ exhibits a relative variability smaller than $50\%$ across the range of time increments $\tau$ used in the calculation.

\section{The energy cascade rate at MHD scales}\label{sec:theory}

Starting from the compressible MHD equations and assuming fully developed, homogeneous turbulence, i.e., ${\rm Re},{\rm Re}_m \to \infty$ and a statistically steady state in which forcing balances dissipation \citep[e.g.,][]{B2018,Ba2020}, one can derive an exact relation valid in the MHD inertial range,
\begin{equation}
-2\varepsilon_c = \frac{1}{2}\,{\boldsymbol\nabla}_\ell \!\cdot \mathbf{F}_c + S_C + S_H + M_\beta,
\label{TermComplete}
\end{equation}
where $\varepsilon_c$ is the total compressible energy cascade rate, $\mathbf{F}_c$ denotes the compressible flux vector, and $S_C$, $S_H$, and $M_\beta$ are the compressible source, hybrid, and plasma-$\beta$--dependent terms, respectively. A complete derivation and explicit expressions for all terms are provided by \citet{A2017b}. In this work we estimate $\varepsilon_c$ using only the flux term $F_c$, since $S_C$, $S_H$ and $M_\beta$ require multi-point measurements of density and Alfv\'en velocity gradients that are not available for the single-spacecraft intervals analysed here \citep{A2019a}. Numerical simulations of compressible MHD turbulence suggest that these additional terms remain subdominant in both subsonic and supersonic regimes \citep[e.g.,][]{A2025,F2020,A2019a}. This indication should nevertheless be taken with caution: simulations cannot reproduce the Reynolds numbers, the $\beta$ range or the expansion of the solar wind plasma, and the relative weight of the source and hybrid terms in the interplanetary medium remains essentially unconstrained observationally.

The validity of the approximation is, moreover, expected to be heliocentric distance-dependent. Compressibility decreases as the wind expands: the turbulent Mach number and the normalized density fluctuation level drop between the near-Sun PSP intervals and 1~au, so that the neglected terms, which are quadratic in the density and Alfv\'en-velocity increments, should be least important at 1~au and most relevant in the innermost PSP orbits \citep{Ch2020,B2023}. Conversely, the expansion itself breaks the assumption of a statistically steady state on which the exact relation rests, and this effect is also stronger close to the Sun. Our near-Sun estimates should therefore be regarded as the least conservative of the sample, and the radial trends reported here as robust in their ordering rather than in their absolute normalization. A quantitative assessment of the neglected terms will require multi-spacecraft configurations such as those provided by MMS or by future constellation missions.

The compressible flux term can be decomposed into two distinct contributions. The first is a fourth-order, Yaglom-like term,
\begin{equation}\label{yaglom}
  \mathbf{F}_{1c} = \Big\langle \Big[\delta(\rho\mathbf{u})\!\cdot\!\delta\mathbf{u} + \delta(\rho\mathbf{u}_A)\!\cdot\!\delta\mathbf{u}_A\Big]\, \delta\mathbf{u} - \Big[\delta(\rho\mathbf{u})\!\cdot\!\delta\mathbf{u}_A + \delta\mathbf{u}\!\cdot\!\delta(\rho\mathbf{u}_A)\Big]\, \delta\mathbf{u}_A \Big\rangle,
\end{equation}
which generalizes the incompressible MHD flux term derived by \citet{P1998b,P1998a}. The second contribution is a purely compressible term,
\begin{equation}\label{com_flux}
  \mathbf{F}_{2c} = 2\,\langle \delta\rho\,\delta e\,\delta\mathbf{u} \rangle,
\end{equation}
where $e=c_s^2\log(\rho/\rho_0)$ \citep{A2017b,S2021} and $c_s$ is the sound speed. Here $\delta\alpha \equiv \alpha' - \alpha$ and $\langle\cdot\rangle$ is a time average over the MHD inertial range. Different thermodynamic closures affect $\mathbf{F}_{2c}$ only through the specific form of $e(\rho,p)$ \citep[see,][]{S2021,B2023}. In the incompressible limit ($\rho \to \rho_0 = \mathrm{const}$), $\mathbf{F}_{1c}$ reduces to the fully developed incompressible MHD flux derived by \citet{P1998b,P1998a},
\begin{equation}\label{eq:fi}
\mathbf{F}_i = \rho_0 \Big\langle \big[(\delta \mathbf{u})^2 + (\delta \mathbf{u}_A)^2\big]\delta \mathbf{u} - 2(\delta \mathbf{u}\!\cdot\!\delta \mathbf{u}_A)\,\delta \mathbf{u}_A \Big\rangle.
\end{equation}
Finally, the cascade rates $\varepsilon_{1c}$, $\varepsilon_{2c}$ and $\varepsilon_i$ correspond to the longitudinal components of Eqs.~(3), (4), and (5), respectively, projected along the solar wind flow direction (via the Taylor hypothesis) and divided by $-4\ell\rho_0/3$.

\section{Results}\label{sec:results}

\begin{figure*}
\centering
\includegraphics[width=0.9\textwidth]{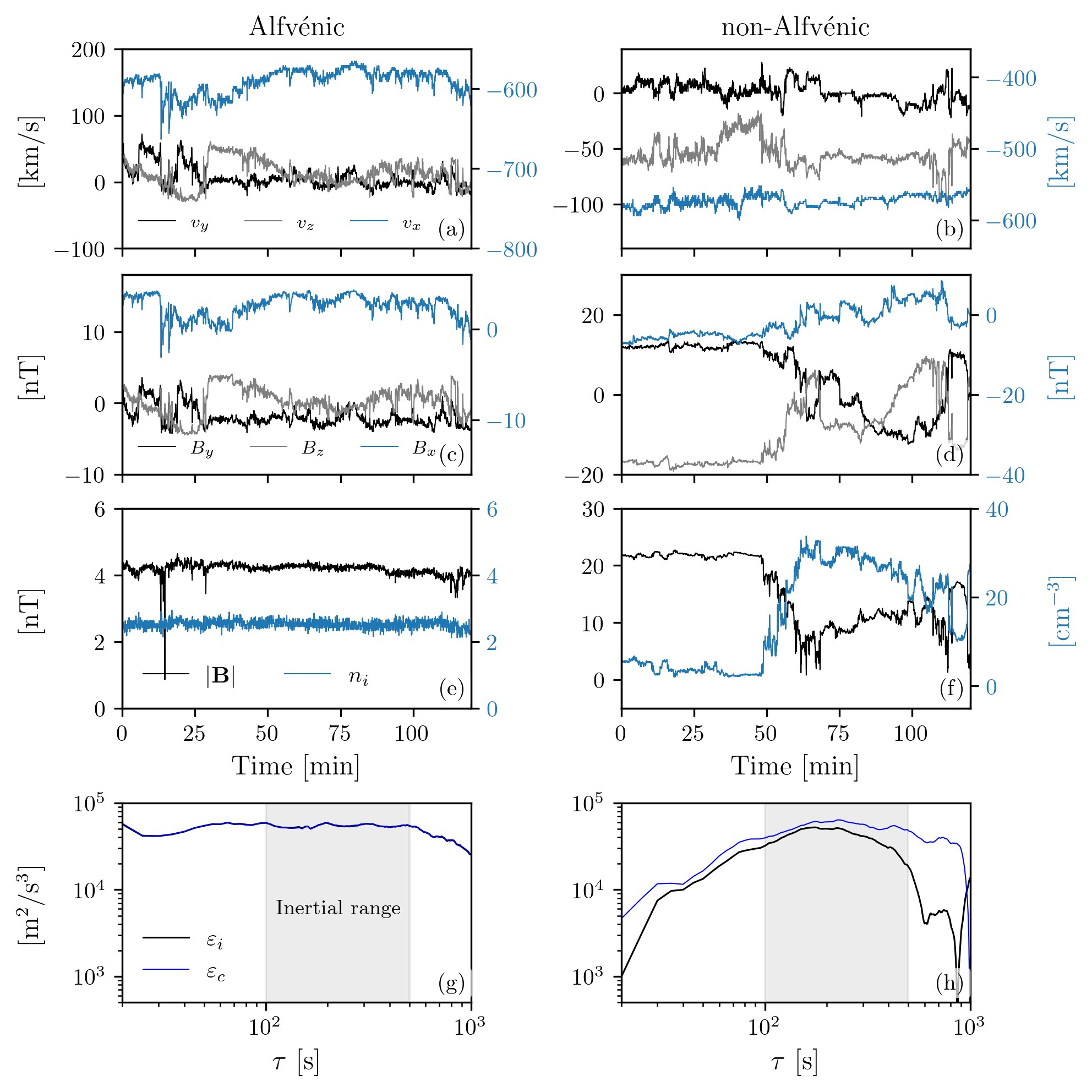}
\caption{Representative 2-hour intervals of Alfvénic (left) and non-Alfvénic (right) solar wind observed by WIND. Panels (a,b): velocity components, with $v_y$ (black) and $v_z$ (gray) sharing the left-hand vertical axis and $v_x$ (blue) referred to the right-hand vertical axis. Panels (c,d): magnetic field components, with $B_y$ (black) and $B_z$ (gray) sharing the left-hand vertical axis and $B_x$ (blue) referred to the right-hand vertical axis. Panels (e,f): magnetic field magnitude $|\mathbf{B}|$ (black) and ion density $n_i$ (blue). Panels (g,h): compressible ($\varepsilon_c$, blue) and incompressible ($\varepsilon_i$, black) energy cascade rates as a function of time-lag $\tau$; the gray shaded region marks the MHD inertial range ($\tau \in [100,500]$~s) used to compute the mean cascade rates throughout this work.}
\label{fig:timeseries}
\end{figure*}

Figure~\ref{fig:timeseries} shows representative 2-hour intervals of Alfvénic (left column) and non-Alfvénic (right column) solar wind observed by WIND. The top three rows display the three components of the velocity (a,b) and magnetic field (c,d), together with the magnetic field magnitude and ion density (e,f). The fields are in Geocentric Solar Ecliptic (GSE) coordinates, where $\hat x$ points sunward, $\hat z$ is perpendicular to the ecliptic plane (positive north), and $\hat y$ completes the right-handed system. The Alfvénic interval is characterized by large-amplitude, highly correlated velocity and magnetic field fluctuations with nearly constant $|\mathbf{B}|$ and low density variability, consistent with the dominance of outward-propagating Alfvén waves. The non-Alfvénic interval, by contrast, exhibits weaker velocity fluctuations, magnetic magnitude variations, and larger density fluctuations, reflecting the slightly more compressible and less coherent character of this wind type. The bottom panels (g,h) show the corresponding compressible ($\varepsilon_c$, blue) and incompressible ($\varepsilon_i$, black) cascade rates as a function of the time lag $\tau$, with the shaded region indicating the MHD inertial range used in the analysis ($\tau \in [100, 500]$~s). In the non-Alfvénic case, $\varepsilon_c > \varepsilon_i$ across the inertial range, while in the Alfvénic case the two estimates are nearly coincident, illustrating the regime dependent behavior that motivates the statistical analysis below.

\subsection{Statistics of Alfvénic and non-Alfvénic fluctuations}

\begin{figure}
\centering
\includegraphics[width=0.9\hsize]{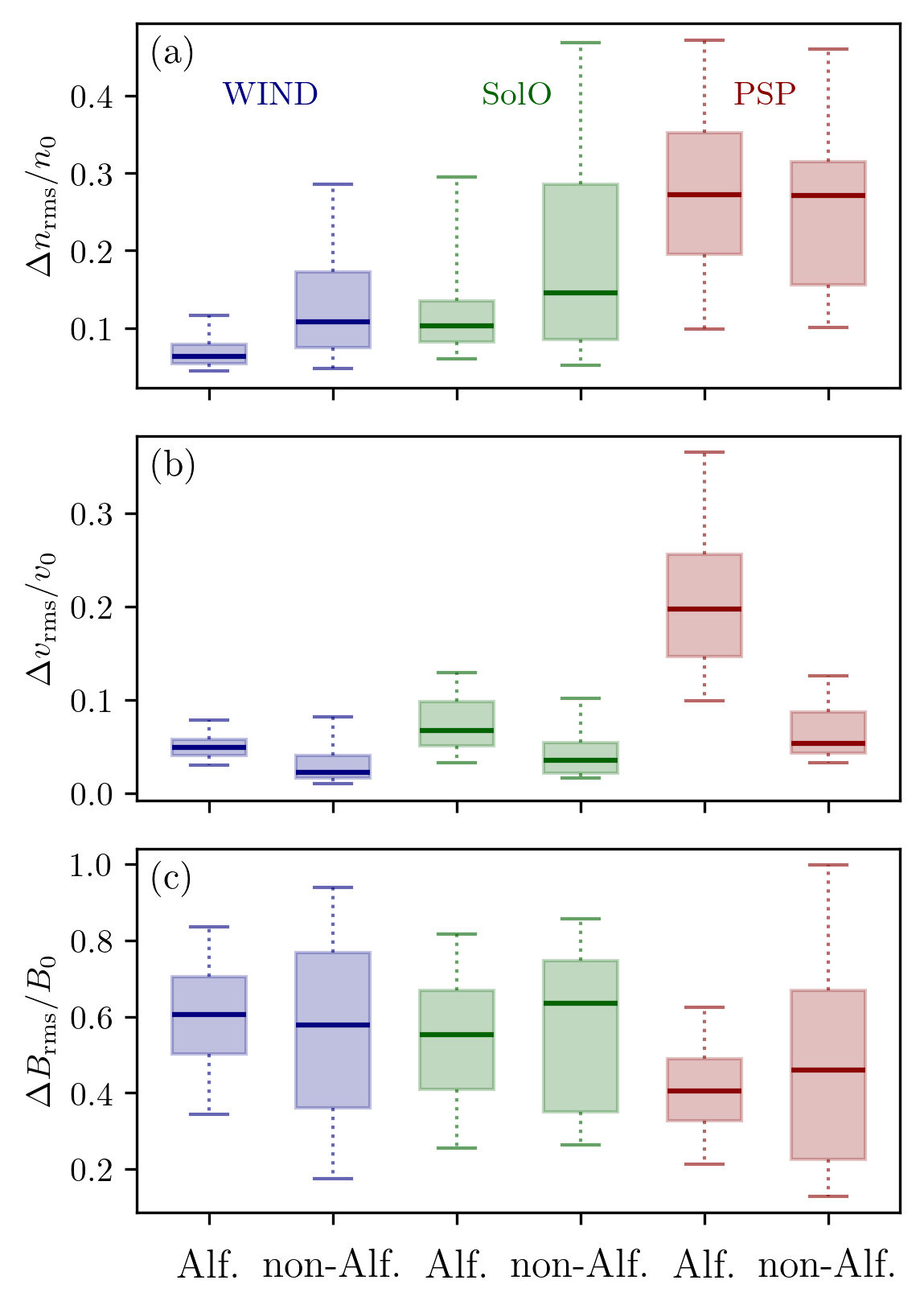}
\caption{Box plots of the normalized RMS fluctuation amplitudes for (a)~density, (b)~velocity, and (c)~magnetic field in Alfvénic and non-Alfvénic solar wind intervals observed by WIND (blue), SolO (green), and PSP (red), respectively. Boxes indicate the inter-quartile range (25th--75th percentile) and the bold horizontal line indicates the median; whiskers extend to the 5th and 95th percentiles.}
\label{fig:rms_boxplots}
\end{figure}

We characterize the level of fluctuation of each quantity using the root-mean-square (RMS), defined as $\Delta x_\mathrm{rms} = {\langle (x - x_0)^2\rangle}^{1/2}$, where the result is normalized to the mean background value $x_0$. This normalization provides a dimensionless measure of the fluctuation level that is directly comparable across different spacecraft observations at different heliocentric distances and solar wind conditions. Figure~\ref{fig:rms_boxplots} shows the box plots of normalized RMS fluctuation amplitudes for (a) density (${\langle(n-n_0)^2\rangle}^{1/2}/n_0$), (b) velocity (${\langle (\mathbf{v}-\mathbf{v}_0)^2 \rangle}^{1/2}/v_0$), and (c) magnetic field (${\langle (\mathbf{B}-\mathbf{B}_0)^2\rangle}^{1/2}/B_0$), computed across all available intervals from WIND (blue), SolO (green) and PSP (red) and separated by type of solar wind (i.e., Alfvénic or non-Alfvénic).

Density fluctuations from WIND and SolO in Alfvénic intervals are systematically lower than in non-Alfvénic intervals, with WIND and SolO medians of $\Delta n_\mathrm{rms}/n_0 \sim 0.064$ and $0.103$ for Alfvénic intervals, compared to $\sim 0.108$ and $\sim 0.146$ for non-Alfvénic intervals, respectively. This result is consistent with the well-established compressive character of non-Alfvénic solar wind: intervals lacking a dominant Alfvénic component tend to exhibit stronger density variations driven by compressive fluctuations and pressure-balanced structures \citep{Tu1995,BC2013,DA2015,R2024b}. A notable exception is PSP Alfvénic intervals, which show significantly larger density fluctuations ($\sim 0.27$) compared to WIND and SolO, likely reflecting the enhanced compressibility of the young solar wind close to the Sun, before significant radial expansion \citep{Ch2020,B2023,zhao2025,Go2026}.

Velocity fluctuations show the opposite ordering: they are markedly larger in Alfvénic intervals (median $\Delta v_\mathrm{rms}/v_0 \sim 0.051$) than in non-Alfvénic intervals ($\sim 0.023$), by a factor of $\sim 2.2$ when all observations are combined. This is a direct signature of the Alfvén wave content: Alfvénic intervals carry large-amplitude incompressible velocity fluctuations nearly in equipartition with magnetic fluctuations, whereas non-Alfvénic intervals are dynamically less variable \citep{B1971,Goldstein1995}. Again, PSP Alfvénic intervals stand out with a median of $\sim 0.20$, approximately 4 times larger than WIND values at 1~au, indicating that the velocity fluctuation amplitude increase substantially with decreasing heliocentric distances \citep{Shi2023,A2022}.

Magnetic field fluctuations show no statistically significant difference between the two classes ($\Delta B_\mathrm{rms}/B_0 \sim 0.59$ for Alfvénic and $\sim 0.57$ for non-Alfvénic, all missions combined). The broad distributions and overlapping inter-quartile ranges indicate that the normalized magnetic fluctuation amplitude alone does not discriminate between Alfvénic and non-Alfvénic wind, and it is controlled mainly by the expansion. Interestingly, PSP shows lower magnetic fluctuation levels ($\sim 0.40$) compared to WIND ($\sim 0.62$) for Alfvénic intervals, in contrast to the radial trend seen in panels (a) and (b), possibly reflecting a different balance between wave amplitudes and background field strength closer to the Sun \citep{BC2013,A2022,Go2026}.

\subsection{Statistics of compressible MHD energy cascade rates}\label{sec:cascade}

\begin{figure}
\centering
\includegraphics[width=0.9\hsize]{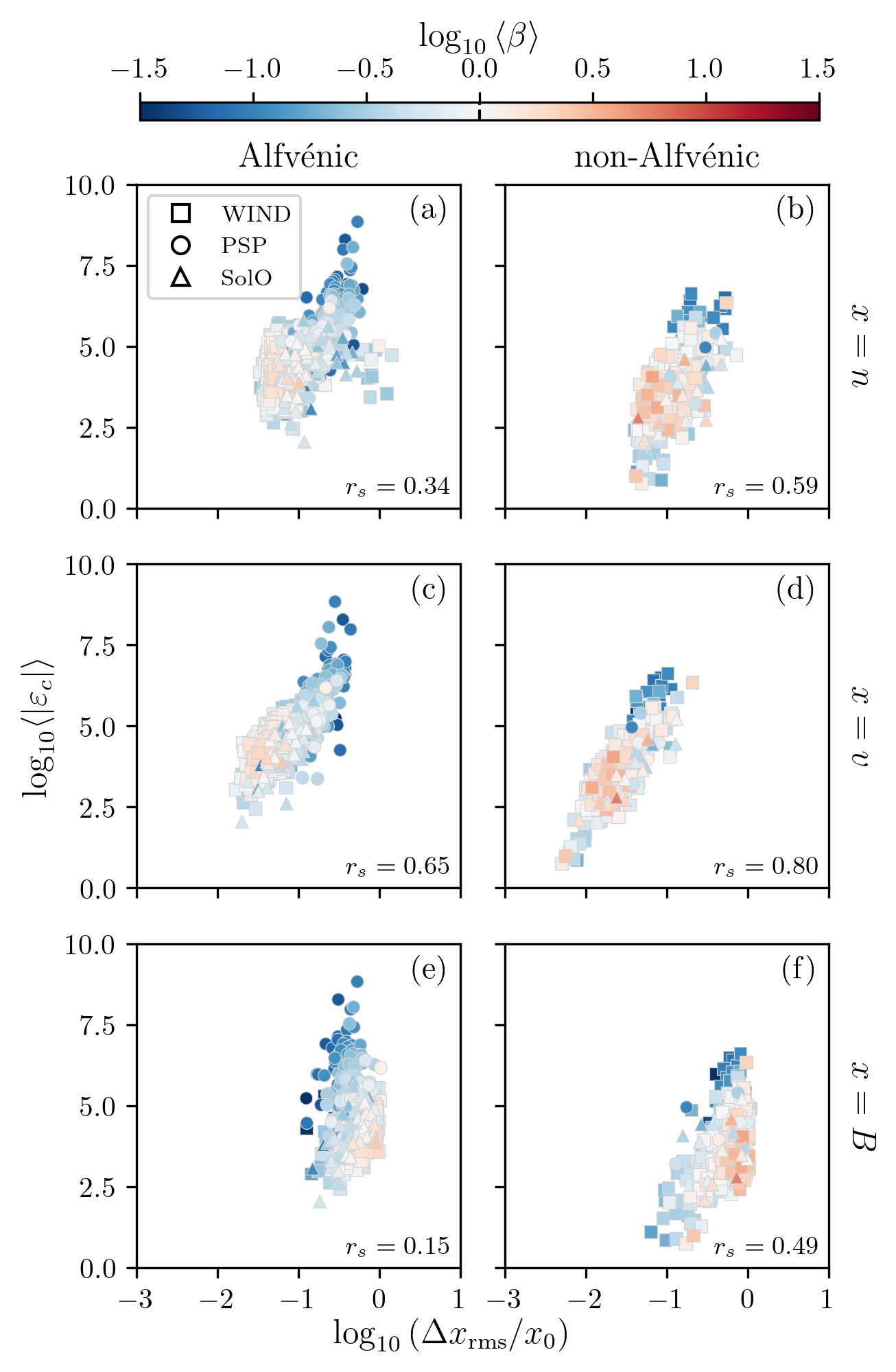}
\caption{Compressible energy cascade rate as a function of the (a-b)~density, (c-d)~velocity, and (e-f)~magnetic field normalized RMS fluctuation amplitudes, under different solar wind regimes. The colorbar corresponds to $\log_{10}\langle\beta\rangle$, where $\langle\beta\rangle$ is the mean plasma beta.}
\label{fig:epsc_vs_fluc_beta}
\end{figure}

Figure~\ref{fig:epsc_vs_fluc_beta} displays the compressible energy cascade rate $\langle|\varepsilon_c|\rangle$ (computed following the isothermal exact law of \citet{A2017b}, for $\tau\in[100,500]$~s) as a function of the normalized RMS fluctuation level $\Delta x_{\mathrm{rms}}/x_0$ for (a-b) density, (c-d) velocity, and (e-f) magnetic field. Events are grouped into Alfv\'enic (left column) and non-Alfv\'enic (right column) types, with marker shapes corresponding to WIND (squares), PSP (circles), and SolO (triangles). Each point is colored by $\log_{10}\langle\beta\rangle$, using a divergent blue--red colormap centered at $\beta=1$ \citep[see, e.g.,][]{V2019}, where $\beta = P_\mathrm{th}/P_\mathrm{mag} = n k_B T_p / (|{\bf B}|^2/8\pi)$ is the ratio of thermal to magnetic pressure and $k_B$ is the Boltzmann constant. To quantify the relationship between fluctuation amplitude and cascade rate, we compute the Spearman rank correlation coefficient $r_s$ \citep{S1904}, a non-parametric measure bounded between $-1$ and $+1$ that captures monotonic associations without assuming linearity. A positive correlation between fluctuation amplitude and compressible cascade rate is observed across all panels, with the strength depending markedly on the fluctuation variable. Velocity fluctuations exhibit the highest relationship ($r_s = 0.65$ for the Alfv\'enic population and $r_s = 0.80$ for the non-Alfv\'enic one), suggesting that this variable is a strong predictor of the compressible energy flux in the explored solar wind regimes. Density fluctuations yield moderate correlations ($r_s = 0.34$ and $0.59$ for Alfv\'enic and non-Alfv\'enic, respectively), while magnetic field fluctuations show the weakest association ($r_s = 0.15$ and $r_s = 0.49$, respectively). This hierarchy suggests that, although density compressibility contributes to the cascade \citep{H2017a,A2021,B2023}, it is the velocity fluctuation amplitude that is most directly related to the magnitude of $\langle|\varepsilon_c|\rangle$.

Substantial differences emerge between spacecraft, reflecting the radial evolution of turbulence in the inner heliosphere \citep[e.g.,][]{R1995,Bavassano2001,Shi2023}. PSP events, sampled at heliocentric distances as close as ${\sim}$~0.05~au, display systematically larger fluctuation levels and cascade rates than WIND at 1~au: for Alfv\'enic intervals, the median $\log_{10}|\varepsilon_c|$ increases from $4.47$ (WIND) to $5.66$ (PSP), i.e., roughly a factor of ${\sim}$~15 enhancement. Concurrently, median density and velocity fluctuation levels are higher by $\sim 0.6$ and $\sim 0.5$~decades, respectively (median $\log_{10}(\Delta n_{\mathrm{rms}}/n_0)=-0.60$ for PSP vs.\ $-1.19$ for WIND; median $\log_{10}(\Delta v_{\mathrm{rms}}/v_0)=-0.74$ vs.\ $-1.31$). SolO events show a distinct pattern: their median cascade rate ($\log_{10}|\varepsilon_c| = 4.23$) falls below WIND despite sampling closer heliocentric distances (${\sim}$~0.3--$1$~au), while their density fluctuation level (median $\log_{10}(\Delta n_\mathrm{rms}/n_0) = -0.97$) lies between WIND ($-1.19$) and PSP ($-0.60$), suggesting that the SolO Alfvénic sample captures a mix of near- and far-Sun intervals whose cascade rates are not yet as elevated as those of the innermost PSP orbits. These trends are in agreement with the radial amplification of the compressible cascade rate previously reported \citep{B2016c,H2017a,A2021,B2023}.

The plasma $\beta$ dependence reveals a clear organization of the scatter plots. In the Alfv\'enic column, most events lie in low-$\beta$ regime: $79.3\%$ for WIND, $82.7\%$ for SolO, and $96.4\%$ for PSP. The PSP cluster appears predominantly blue, indicating that these near-Sun Alfv\'enic intervals are strongly magnetically dominated, consistent with the expected decrease of $\beta$ toward the Sun \citep[e.g.,][]{M2011,V2019}. In contrast, the non-Alfv\'enic population exhibits a broader $\beta$ distribution, with only $73.7\%$ (WIND), $52.9\%$ (SolO), and $83.3\%$ (PSP) of events below $\beta=1$. This is consistent with the association of non-Alfv\'enic wind with slower, denser, and thermally more pressure-balanced streams \citep{DA2015,BC2013}. Non-Alfv\'enic intervals also exhibit systematically lower cascade rates than their Alfv\'enic counterparts (median $\log_{10}|\varepsilon_c| = 3.64$ vs. $4.51$ for the combined dataset), together with weaker velocity fluctuations (median $\log_{10}(\Delta v_{\mathrm{rms}}/v_0) = -1.65$ vs. $-1.30$).

The lower values of $\langle|\varepsilon_c|\rangle$ found in the non-Alfv\'enic population deserve a careful comparison with previous results. \citet{Ma2012}, analysing Ulysses observations of the high-latitude solar wind, reported the opposite behaviour: states with high cross helicity were associated with a reduction of the energy transfer rate, and the slow streams of that dataset displayed larger cascade rates than the neighboring fast streams, an enhancement attributed to the decorrelation of velocity and magnetic fields and hence to a lower degree of Alfv\'enicity \citep{Ma2011}. Consistently, \citet{smith2009} and \citet{St2010} showed that for $|\sigma_c| \gtrsim 0.75$, precisely the threshold adopted here to define Alfv\'enic intervals, the cascade rate of the dominant outward-propagating component can become negative, indicating a back-transfer of energy from small to large scales, a mechanism that would suppress the net forward cascade in the most strongly Alfv\'enic intervals \citep[see also,][]{P2011,MSV2023}.

Two considerations are relevant when comparing these results with ours. First, the present analysis is based on the absolute value $\langle|\varepsilon_c|\rangle$ averaged over the inertial range, so that intervals undergoing an inverse transfer contribute with large positive magnitudes, whereas the quantities discussed by the works above are signed transfer rates. Our result therefore concerns the magnitude of the energy flux and does not directly contradict the reduction of the signed transfer rate at high cross helicity. A related effect was reported by \citet{St2010}, who found that the range of $\sigma_c$ values associated with negative cascade rates broadens as the total energy of the fluctuations decreases; \citet{R2024a} invoked this mechanism to account for the negative median cascade rates measured upstream of Mars with MAVEN, where the total fluctuation energy is comparatively low. The sign of the transfer is therefore not determined by the Alfvénic state of the plasma in isolation. Second, in our sample Alfv\'enicity is not independent of heliocentric distance: the near-Sun PSP intervals, which display the largest cascade rates, are almost exclusively Alfv\'enic, whereas the non-Alfv\'enic population is drawn mostly from WIND at 1~au. Accordingly, the lower $\langle|\varepsilon_c|\rangle$ observed in the non-Alfv\'enic population should not be read as a reduction of turbulent activity, but as a distinct dynamical regime in which density fluctuations and compressible effects provide alternative channels for energy transfer, enabled by the relaxation of the Alfv\'enic constraint on nonlinear interactions (see Section~\ref{sec:decomp}).

\begin{figure*}
\centering
\includegraphics[width=0.75\hsize]{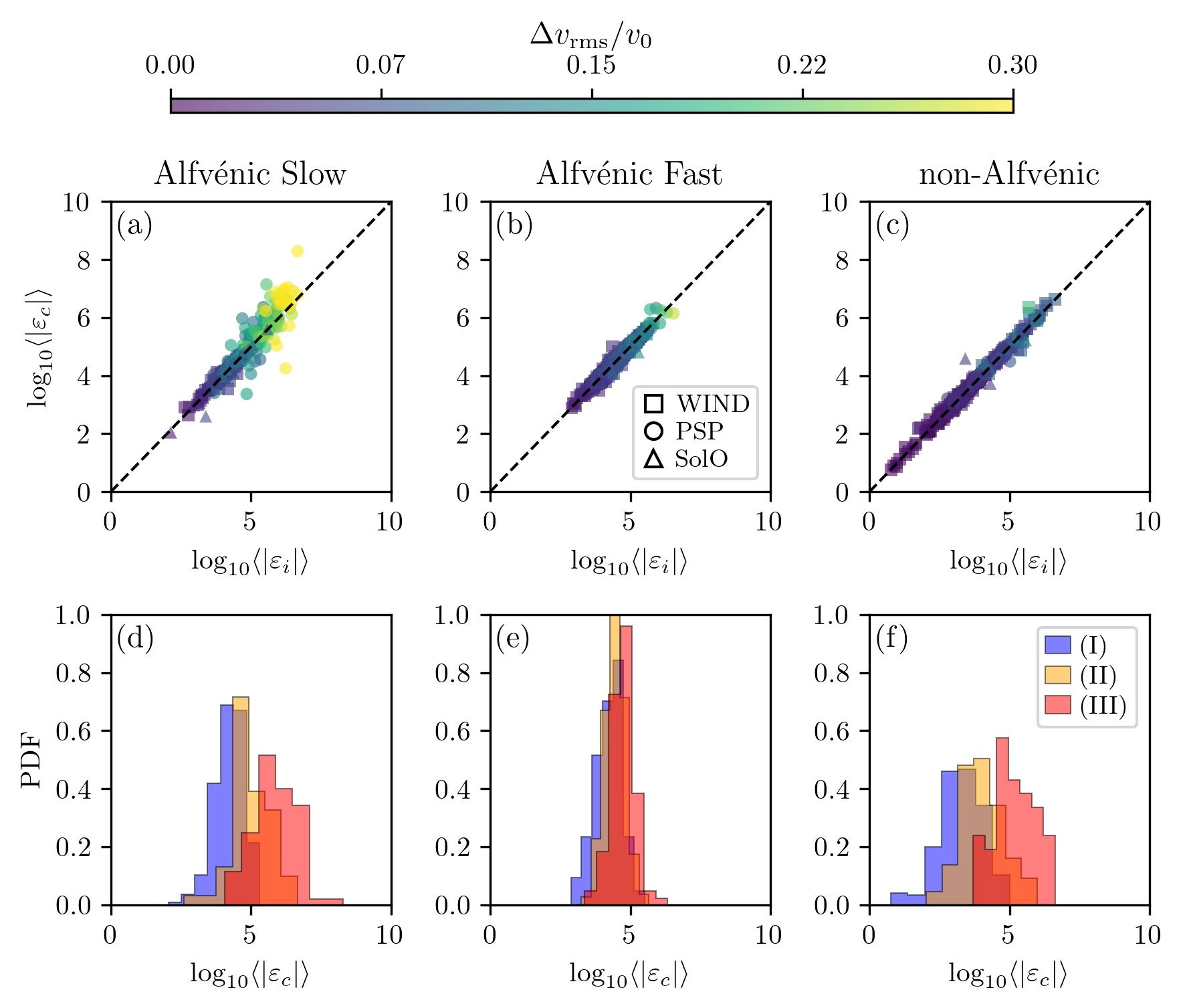}
\caption{Logarithmic relationship between the mean compressible energy cascade rate $\langle|\varepsilon_c|\rangle$ and the mean incompressible energy cascade rate $\langle|\varepsilon_i|\rangle$ at the MHD scales, and the associated mean proton temperature, for three solar wind regimes: Alfvénic slow solar wind ($v_0 \le 400$~km/s), Alfvénic fast solar wind ($v_0 > 500$~km/s), and (slow) non-Alfvénic wind. In the top row (a-c), each point represents an interval colored by the normalized RMS velocity fluctuation level. The bottom row (d-f) shows distributions of $\log_{10}\langle|\varepsilon_c|\rangle$ for intervals grouped into the three proton temperature categories defined in the text, illustrating the higher cascade rates found in the hotter intervals.}
\label{fig:epsi_epsc}
\end{figure*}

Figure~\ref{fig:epsi_epsc} shows the logarithmic relationship between the mean compressible $\langle|\varepsilon_c|\rangle$ and incompressible $\langle|\varepsilon_i|\rangle$ energy cascade rates at the MHD scales, and its dependence on the mean proton temperature ($\langle T_p \rangle$), for three solar wind regimes: Alfv\'enic slow ($v_0 \leq 400$~km/s, $N=372$), Alfv\'enic fast ($v_0 > 500$~km/s, $N=1397$), and non-Alfv\'enic wind ($N=367$). In the top row (a-c), each point represents a turbulent interval colored by the normalized RMS velocity fluctuation level $\Delta v_{\mathrm{rms}}/v_0$. The bottom row (d-f) shows distributions of $\log_{10}\langle|\varepsilon_c|\rangle$ split into three proton temperature categories: I: $T_p \in [1, 13]$~eV (blue), II: $T_p \in [13, 25]$~eV (orange), and III: $T_p \in [25, 100]$~eV (red). The top panels reveal a high correlation between $\langle|\varepsilon_c|\rangle$ and $\langle|\varepsilon_i|\rangle$ across all solar wind regimes. Specifically, Spearman rank coefficients are $r_s = 0.96$ for the Alfv\'enic slow wind, $r_s = 0.99$ for the Alfv\'enic fast wind, and $r_s = 0.99$ for the non-Alfv\'enic population \citep{H2017a,A2021,B2023,J2025}. Nevertheless, a systematic deviation from the dashed diagonal line is clear in the Alfv\'enic slow wind, where $65.9\%$ of intervals satisfy $\langle|\varepsilon_c|\rangle > \langle|\varepsilon_i|\rangle$, a value that increases to $71.3\%$ for the hottest intervals. In contrast, the Alfv\'enic fast solar wind shows no systematic departure from equality, with a fraction above the dashed line close to $50.0\%$ across all temperature bins. The non-Alfv\'enic wind exhibits an intermediate behavior, with $56.1\%$ of events above the dashed diagonal.

Alfvénic slow wind intervals, dominated by PSP near-Sun observations, display the largest velocity fluctuation levels (median $\Delta v_{\mathrm{rms}}/v_0 \sim 0.18$, 5th--95th percentile: $[0.10, 0.35]$). By contrast, Alfvénic fast wind at 1~au, mainly WIND observations, maintains considerably lower normalized velocity amplitudes (median $\Delta v_{\mathrm{rms}}/v_0 \sim 0.05$, 5th--95th percentile: $[0.03, 0.08]$) and is associated with lower proton temperature and density fluctuations, consistent with the quasi-incompressible nature of fast Alfv\'enic streams at larger heliocentric distances \citep{BC2013}, which explains its tighter clustering along the diagonal. Notably, however, the displacement above the diagonal cannot be attributed to velocity fluctuations alone: the compressible formulation explicitly couples to density perturbations through internal energy terms, and it is the intervals with simultaneously elevated density fluctuations ($\Delta n_{\mathrm{rms}}/n_0 > 0.15$, prevalent in PSP near-Sun events) that yield the strongest enhancement of the compressible over the incompressible cascade rate. Thus, large velocity amplitudes and enhanced density fluctuations co-occur in young, near-Sun Alfv\'enic slow wind, jointly driving the compressible energy cascade.

The bottom row histograms demonstrate a clear positive relationship between $\langle|\varepsilon_c|\rangle$ and $T_p$: intervals with higher cascade rates are associated with higher proton temperatures, consistent with the turbulent cascade supplying the energy that is ultimately dissipated into proton heating \citep[e.g.,][]{V1995,C2009,O2013,A2022}. This trend is most pronounced in the Alfvénic slow wind: the median $\log_{10}\langle|\varepsilon_c|\rangle$ increases from $4.24$ (bin~I, $61.8\%$ of events) to $4.89$ (bin~II, $14.2\%$) and $5.77$ (bin~III, $23.4\%$), a rise of nearly $1.5$~decades, with $r_s = 0.72$ between $\log_{10}|\langle\varepsilon_c|\rangle$ and $\log_{10}T_p$. This strong dependence is physically motivated by the internal energy terms in the compressible exact law \citep{A2017b,S2022}. In the Alfvénic fast wind the effect is more moderate: the median $\log_{10}|\langle\varepsilon_c|\rangle$ shifts from $4.31$ (I, $8.2\%$) to $4.47$ (II, $62.9\%$) and $4.74$ (III, $28.8\%$), a total of ${\sim}0.4$~decades ($r_s = 0.40$), reflecting its thermally homogeneous distribution (median $T_p = 20.7$~eV). The non-Alfvénic wind shows a strong coupling ($r_s = 0.70$), with $\log_{10}|\langle\varepsilon_c|\rangle$ rising from $3.28$ (I, $65.4\%$) to $5.22$ (III, $13.6\%$), a nearly two-decade increase consistent with its broad, heterogeneous temperature distribution \citep{DA2015}. Finally, across all regimes, the intervals with the largest compressible cascade rates are consistently the hottest ones, supporting the interpretation that compressible effects, captured through density fluctuations and internal energy terms in the exact law \citep{A2017b,S2022}, play a significant role in amplifying energy transfer rates, particularly in the slow and near-Sun solar wind \citep{C2009,A2021,B2023}.

\subsection{Compressible cascade decomposition}\label{sec:decomp}

\begin{figure*}
\centering
\includegraphics[width=0.95\textwidth]{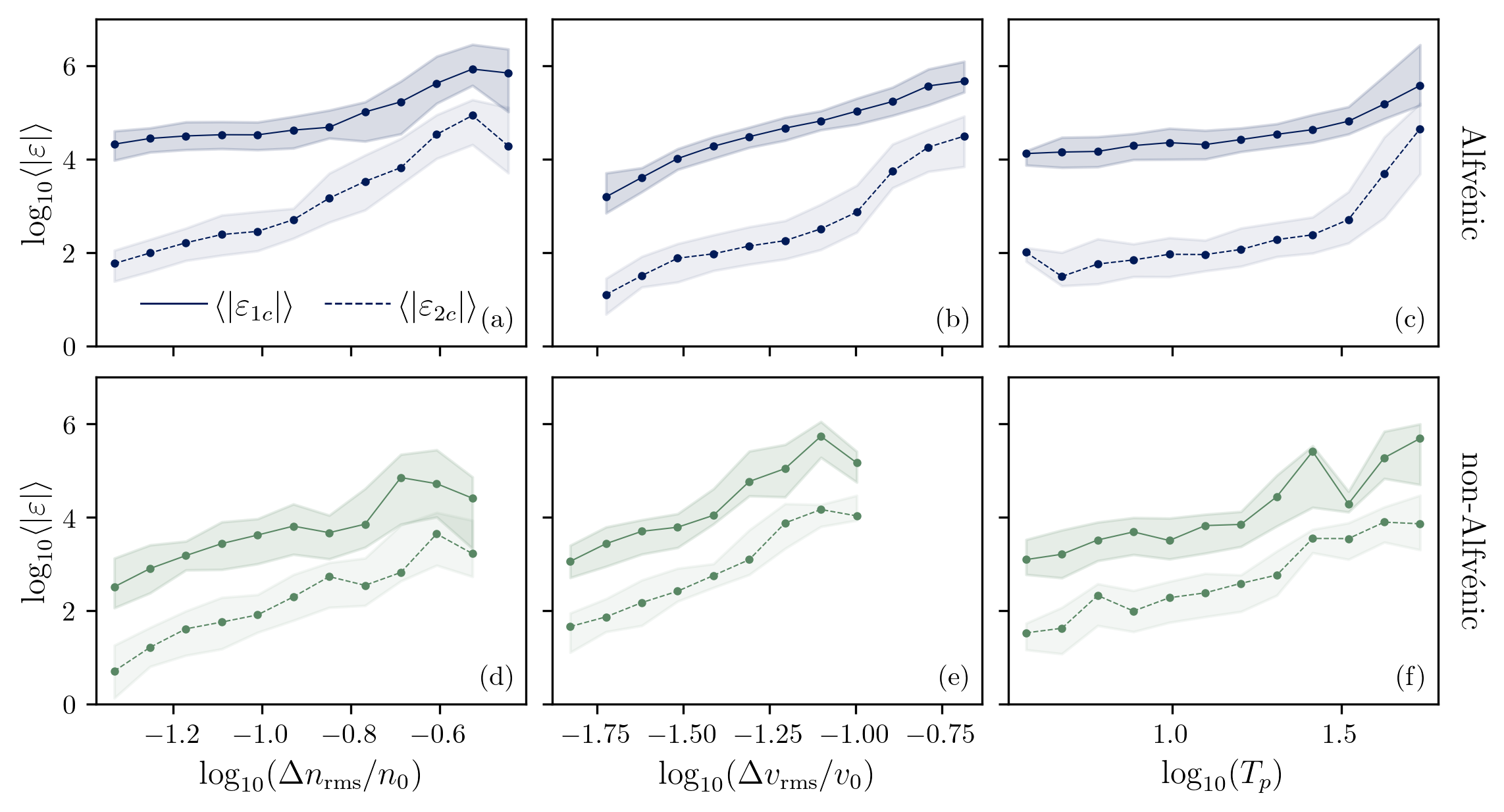}
\caption{Running medians of the two compressible flux components, $\langle|\varepsilon_{1c}|\rangle$ (solid) and $\langle|\varepsilon_{2c}|\rangle$ (dashed), as a function of the normalized density fluctuation amplitude $\Delta n_\mathrm{rms}/n_0$ (a-b), velocity fluctuation amplitude $\Delta v_\mathrm{rms}/v_0$ (c-d), and proton temperature $T_p$ (e-f), for Alfvénic (blue, left) and non-Alfvénic (green, right) solar wind. Shaded bands indicate the interquartile range.}
\label{fig:decomp}
\end{figure*}

To further characterize the physical origin of the compressible cascade, we decompose the total compressible flux $\mathbf{F}_c = \mathbf{F}_{1c} + \mathbf{F}_{2c}$ into its two contributions (Eqs.\,\ref{yaglom} and~\ref{com_flux}) and examine how each cascade component varies with plasma fluctuation amplitude and proton temperature. Figure~\ref{fig:decomp} shows the running medians and interquartile ranges of $\langle|\varepsilon_{1c}|\rangle$ (solid lines) and $\langle|\varepsilon_{2c}|\rangle$ (dashed lines) as a function of $\log_{10}(\Delta n_\mathrm{rms}/n_0)$ (a-b), $\log_{10}(\Delta v_\mathrm{rms}/v_0)$ (c-d), and $\log_{10}(T_p)$ (e-f), separated into Alfvénic (blue, left column) and non-Alfvénic (green, right column) populations.

Across both populations and all three plasma variables, $\langle|\varepsilon_{1c}|\rangle$ systematically exceeds $\langle|\varepsilon_{2c}|\rangle$ by approximately two orders of magnitude, establishing the Yaglom-like term as the dominant channel of compressible energy transfer at MHD scales. This is consistent with numerical results in subsonic compressible MHD turbulence \citep{A2019a,F2020}, where the purely compressible term $\mathbf{F}_{2c}$, which couples density fluctuations to the internal energy increment $\delta e$, remains subdominant unless the turbulent Mach number is large \citep{A2025}. Nonetheless, $\langle|\varepsilon_{2c}|\rangle$ is not negligible: it rises steeply with density fluctuation amplitude (top row), consistent with the direct coupling of $\mathbf{F}_{2c}$ to $\delta\rho$ in Eq.\,\eqref{com_flux}, and is relatively more prominent in the non-Alfvénic population where density variability is enhanced.

The dependence on velocity fluctuation amplitude (middle row) reveals that both components increase monotonically with $\Delta v_\mathrm{rms}/v_0$, a trend confirmed by Spearman rank correlations: $r_s = 0.65$ for $\langle|\varepsilon_{1c}|\rangle$ and $r_s = 0.47$ for $\langle|\varepsilon_{2c}|\rangle$ in the Alfvénic population, with the stronger correlation for the Yaglom-like term supporting the conclusion that the kinematic contribution dominates the cascade in this regime. In the non-Alfvénic population, however, both components correlate much more strongly with $\Delta v_\mathrm{rms}/v_0$ ($r_s = 0.79$ and $r_s = 0.80$ for $\langle|\varepsilon_{1c}|\rangle$ and $\langle|\varepsilon_{2c}|\rangle$, respectively), with $\langle|\varepsilon_{2c}|\rangle$ slightly exceeding $\langle|\varepsilon_{1c}|\rangle$ in correlation strength. This similarity is consistent with the larger ratio $\langle|\varepsilon_{2c}|\rangle / \langle|\varepsilon_{1c}|\rangle$ observed in the non-Alfvénic regime, where the Alfvénic constraint linking density and velocity fluctuations breaks down, allowing compressive fluctuations to drive the $\mathbf{F}_{2c}$ channel as efficiently as the kinematic one \citep{A2017b,B2023}.

The bottom row in Figure~\ref{fig:decomp} shows the two compressible components as a function of the proton temperature. The coupling between $\langle|\varepsilon_{1c}|\rangle$ and $T_p$ is comparatively weak, though statistically robust, in the Alfvénic population ($r_s = 0.50$), consistent with the visually flat trend in panel~(e), while it strengthens considerably in the non-Alfvénic case ($r_s = 0.70$). By contrast, $\langle|\varepsilon_{2c}|\rangle$ correlates with temperature at a level comparable to that of $\langle|\varepsilon_{1c}|\rangle$ in both populations ($r_s = 0.47$ in the Alfvénic wind and $r_s = 0.72$ in the non-Alfvénic wind), and in the non-Alfvénic case this is the strongest correlation with $T_p$ among the four quantities examined. Two distinct, and not mutually exclusive, readings of this association are possible. On the one hand, following the interpretation generally adopted in the literature, the turbulent cascade supplies the energy that is ultimately dissipated into proton heating, so that the intervals with larger $\langle|\varepsilon_{2c}|\rangle$ are expected to be the hotter ones \citep[e.g.,][]{V1995,C2009,O2013,A2022}. On the other hand, $\mathbf{F}_{2c}$ involves the internal energy increment $\delta e \propto c_s^2\,\delta\rho/\rho_0$, so that a hotter, more compressible plasma mechanically enhances this particular channel. The present observations cannot disentangle the two effects, but both act in the same direction and are consistent with the steeper temperature dependence found for $\langle|\varepsilon_{2c}|\rangle$. These results provide a direct, quantitative observational confirmation that the cascade--temperature coupling discussed previously is primarily driven by $\mathbf{F}_{2c}$, while the overall magnitude of the compressible cascade remains controlled by the Yaglom-like term $\mathbf{F}_{1c}$, whose sensitivity to $T_p$ only becomes appreciable once the Alfvénic constraint is relaxed.

\section{Discussion and conclusions}\label{sec:disc}

The systematic reported differences in density, velocity, and magnetic fluctuation amplitudes between Alfv\'enic and non-Alfv\'enic intervals at different heliocentric distances are fully consistent with, and extend, the observational picture established over the past decades \citep[e.g.,][]{Shi2021,K2021,Go2026}. Our multi-spacecraft statistics confirm this at a quantitative level: velocity fluctuations are roughly a factor of 2.2 larger in Alfv\'enic intervals relative to non-Alfv\'enic ones, directly reflecting the presence of large-amplitude, Alfv\'en like fluctuations in near equipartition with the magnetic field \citep{B1971, Goldstein1995}. Conversely, the lower density fluctuations in Alfv\'enic intervals are also expected by the nearly incompressible character of Alfv\'enic turbulence: for nearly incompressible, typically fast Alfv\'enic solar wind, density fluctuations are mostly passively advected by magnetic and velocity fields, which dominate the dynamics, whereas in the more compressible, typically slow solar wind, density fluctuations actively contribute to the nonlinear cascade \citep{K2021, H2017a,Br2014}. The exception we observe for PSP Alfv\'enic intervals, which show significantly larger density fluctuations than WIND or SolO, is consistent with recent multi-mission characterizations of the inner heliosphere, where solar wind compressibility is likely affected by both expansion effects and compressible dynamics governed by local plasma conditions \citep{Go2026}. The absence of a statistically significant difference in normalized magnetic field fluctuation amplitudes between Alfvénic and non-Alfvénic populations is consistent with the finding that, while both wind streams carry different total turbulent energies, $\Delta B_\mathrm{rms}/B_0$ is a quantity that depends on heliocentric distance and plasma $\beta$ rather than on the degree of velocity-magnetic field alignment \citep{BC2013,DA2021,DA2022}. This contrasts with velocity fluctuations, which are directly tied to Alfvénic wave content through the $\mathbf{u}$-$\mathbf{b}$ correlation: intervals with high cross-helicity carry large-amplitude, incompressible velocity fluctuations in near-equipartition with the magnetic field \citep{B1971, Goldstein1995}, whereas non-Alfvénic intervals lack this coherent wave-driven component, resulting in systematically weaker velocity fluctuations despite comparable normalized magnetic amplitudes.

The hierarchy of correlations between individual fluctuation amplitudes and the compressible cascade rate $\langle|\varepsilon_c|\rangle$ constitutes one of the central results of this work: velocity fluctuations yield the tightest coupling ($r_s \sim 0.65$--$0.80$), followed by density ($r_s \sim 0.34$--$0.59$), with magnetic fluctuations showing the weakest link ($r_s \sim 0.15$--$0.49$). Our finding that $\langle|\varepsilon_c|\rangle$ correlates more strongly with density fluctuations in non-Alfv\'enic wind than in Alfv\'enic wind ($r_s = 0.59$ vs.\ $0.34$) are in agreement with \citet{H2017a}, who found that the energy cascade rate is amplified particularly in the slow solar wind, exhibiting a power-law scaling with the turbulent Mach number and a lower level of spatial anisotropy. This amplification arises because density fluctuations enter the compressible exact law both through the generalized Yaglom flux and through the purely compressible terms, which couple density to the velocity field; these compressible contributions become quantitatively more significant in non-Alfv\'enic wind, where the Alfv\'enic constraint on density variability is relaxed \citep{A2017b, B2016c}. In this regime, density fluctuations can substantially amplify the turbulent cascade rate with respect to the incompressible model \citep{B2023, A2021}, and our Spearman rank analysis shows that this effect is consistently stronger in the non-Alfv\'enic population. The decomposition of the compressible energy cascade into its two components (Fig.~\ref{fig:decomp}) confirms this picture directly: the purely compressible term $\langle|\varepsilon_{2c}|\rangle$ shows a markedly steeper dependence on density fluctuation amplitude in the non-Alfvénic regime, while the Yaglom-like term $\langle|\varepsilon_{1c}|\rangle$ remains the dominant contribution across both populations. The weak correlation of magnetic fluctuations with $\langle|\varepsilon_c|\rangle$ in Alfv\'enic intervals ($r_s = 0.15$) is also physically natural: since both Alfv\'enic and non-Alfv\'enic populations display similarly broad and overlapping distributions of $\Delta B_\mathrm{rms}/B_0$, the magnetic fluctuation amplitude does not affect the energy flux differently within the two regimes. This is consistent with the known dominance of transverse, incompressible fluctuations over compressive ones across most of the inertial range \citep{Tu1995, BC2013,A2018b}.

The near-perfect correlations between $\langle|\varepsilon_c|\rangle$ and $\langle|\varepsilon_i|\rangle$ across all three solar wind regimes demonstrate that the compressible and incompressible cascades are tightly coupled at MHD scales, extending and reinforcing previous numerical and observational findings \citep{H2017a,A2021,S2021,B2023,A2025}. Our analysis adds the important distinction that systematic departures from $\varepsilon_c = \varepsilon_i$ are organized by the type of wind: Alfvénic slow wind shows a systematic excess of $\langle|\varepsilon_c|\rangle$ over $\langle|\varepsilon_i|\rangle$ ($65.9\%$ of events above the diagonal, rising to $71.3\%$ for the hottest temperature bin), while Alfvénic fast wind clusters tightly along the diagonal and the non-Alfvénic population shows an intermediate behavior ($56.1\%$ above the diagonal). This organization is consistent with the role of internal energy contributions in the compressible exact law \citep{A2017b,S2022}: in Alfvénic slow wind, where proton temperatures are highest and density fluctuations are comparatively more active than in fast wind, the thermal terms provide an additional positive flux that systematically shifts $\langle|\varepsilon_c|\rangle$ above $\langle|\varepsilon_i|\rangle$. At the same time, our results show that velocity fluctuation amplitude is the primary control on $\langle|\varepsilon_c|\rangle$ across all regimes, suggesting that kinematic and thermodynamic compressibility play distinct but complementary roles in regulating the compressible cascade. The non-Alfv\'enic population shows a small excess of $\langle|\varepsilon_c|\rangle$ over $\langle|\varepsilon_i|\rangle$, consistent with the findings of \citet{B2023}, who reported that compressible and incompressible cascade rates increase together toward the Sun, with the compressible estimate moderately exceeding the incompressible one as compressibility in the plasma increases. The stronger correlations observed in the non-Alfv\'enic wind despite its statistically slightly lower absolute cascade rates may reflect a more direct compressible coupling in the absence of dominant Alfv\'enic fluctuations, in line with the interpretation that compressible fluctuations become dynamically active and sensitive to local structures when the Alfv\'enic constraint is broken \citep[e.g.,][]{SV2007, Br2014}.

The strong positive correlation between the energy cascade rate in the inertial range and the proton temperature is firmly based in both theory and prior observational evidence \citep{V1995,C2009,St2009,K2013,Ba2020,B2020,A2021,A2022,S2022,B2023}. The incompressible cascade rate was already known to correlate with proton temperature and kinetic effects \citep{O2013,A2022}. Specifically, enhanced cascade rates are associated with proton temperature anisotropy and plasma heating, with protons found to be several times hotter in high-cascade-rate intervals \citep{V1995,C2009,A2021}. Our results show that the analogous relationship holds, and is in fact stronger in several regimes, for the compressible cascade, which explicitly incorporates density and velocity fluctuations and internal-energy contributions in the exact law. The steepest variation of the cascade rate with proton temperature appears in the Alfvénic slow wind (a nearly 1.5 decade increase from cool to hot events) and in the non-Alfv\'enic wind (a nearly two-decade increase), while the Alfv\'enic fast wind shows a weaker dependence owing to its thermally more homogeneous distribution. This contrast is consistent with the established result that Alfv\'enic periods occur mainly when density fluctuations are low and temperature is relatively high, while the broad temperature distributions of the slow and non-Alfv\'enic streams \citep{DA2015} provide the range needed to reveal the cascade--temperature coupling. The radial evolution enters here as well: the systematically higher cascade rates and velocity fluctuation levels in PSP events, compared to WIND or SolO, reflect the well-documented amplification of turbulent activity closer to the Sun, as reported by multi-spacecraft studies using exact law approaches \citep[e.g.,][]{B2016c, H2017a, A2020, R2022, R2024a, A2021, B2023}. The joint effect of elevated density and velocity fluctuations in near-Sun Alfv\'enic intervals, particularly the PSP dominated slow wind, thus drives the compressible cascade above the incompressible estimate, an excess reinforced by the higher proton temperatures of these intervals through the internal energy terms of the exact law, whose amplitude scales with $c_s^2$, providing a coherent observational picture in which both flux terms in the exact relation act together to modulate energy transfer in the young solar wind, and pointing toward the need for compressible, multi-fluid descriptions to fully capture the dissipation budget at MHD scales \citep{A2018b,Ba2020, S2022}. We note, however, that the present observations cannot establish the direction of causality between the cascade rate and the proton temperature. The cascade decomposition in Fig.~\ref{fig:decomp} further shows that, while both compressible components correlate significantly with $T_p$, the relative variation with temperature is markedly larger for $\langle|\varepsilon_{2c}|\rangle$, which rises by roughly two orders of magnitude over the sampled $T_p$ range compared to the comparatively flat plateau of $\langle|\varepsilon_{1c}|\rangle$ in the Alfvénic population. This asymmetry is consistent with the purely compressible term's direct dependence on the internal energy increment $\delta e$, which is thermodynamically sensitive, while $\langle|\varepsilon_{1c}|\rangle$ is set primarily by the kinematic (velocity driven) cascade.

Finally, we recall that the lower values of $\langle|\varepsilon_c|\rangle$ found in the non-Alfv\'enic population (Section~\ref{sec:cascade}) should be interpreted with care, since the present analysis is based on the absolute value of the cascade rate and a sign-resolved comparison with the results of \citet{Ma2011,Ma2012} is left for future work. Beyond this, further limitations of the present study should be acknowledged. First, our analysis is restricted to the MHD range of turbulence, thereby neglecting potentially important dissipation channels captured by more advanced descriptions, such as Hall MHD, CGL MHD, or full two-fluid models \citep{Ba2020,S2022,A2018b,A2019a,Hu2019}. Second, the turbulence may not always be in a fully developed state, particularly in non-Alfvénic regimes, where intermittency and structural complexity are enhanced. In such cases, the use of local or scale-dependent energy cascade rates could provide a more accurate characterization of the energy transfer. A natural extension of this work is therefore to perform a scale-by-scale analysis of turbulence within a coarse-grained framework for plasma dynamics \citep{M2022,M2023}.

\begin{acknowledgements}
We acknowledge the WIND, Solar Orbiter and Parker Solar Probe instrument teams for making the data publicly available. N.A. acknowledges financial support from the following grants: ECOS SUD 2022 \#A22U02 CNRS/CONICET, PIP Grant No.~11220200101752, UBACyT Grant No.~20020220300122BA and 20020250100107BA and Redes de Alto Impacto REMATE from Argentina. C.A.G. is supported by NSF SHINE award number 2400967. N.R. is supported by NASA under award number 80GSFC24M0006 and by the NASA Solar Orbiter Project Science.
\end{acknowledgements}

\section*{Data availability}
Parker Solar Probe, Solar Orbiter, and WIND data are publicly available from NASA CDAWeb: \url{https://cdaweb.gsfc.nasa.gov}.

\end{document}